\documentclass[prd,aps,preprint,tightenlines,nofootinbib,superscriptaddress]{revtex4}

\usepackage{epsfig}
\usepackage{amsmath}
\input{epsf}
\usepackage{psfrag}

\usepackage[usenames,dvipsnames]{color}

\newcommand{\nn}{\nonumber \\}
\newcommand{\beq}{\begin{eqnarray}}
\newcommand{\eeq}{\end{eqnarray}}
\newcommand{\Slash}[1]{{\ooalign{\hfil/\hfil\crcr$#1$}}}

\begin{document}

\vspace*{2cm}
\title{Spin-dependent Pomeron and Odderon in elastic proton-proton scattering}

\author{Yoshikazu Hagiwara}
\affiliation{ Key laboratory of Particle Physics and Particle Irradiation (MOE), \\
Institute of frontier and interdisciplinary science,\\
Shandong University (QingDao), Shandong 266237, China}

\author{Yoshitaka Hatta}
\affiliation{ Physics Department, Brookhaven National Laboratory, Upton NY 11973,  USA}

\author{Roman Pasechnik}
\affiliation{Department of Astronomy and Theoretical Physics, Lund University, 221 00 Lund, Sweden  \vspace*{1cm}}

\author{Jian Zhou}
\affiliation{ Key laboratory of Particle Physics and Particle Irradiation (MOE), \\
Institute of frontier and interdisciplinary science,\\
Shandong University (QingDao), Shandong 266237, China}

\begin{abstract}
\vspace*{0.5cm}
We introduce a new model of near-forward elastic proton-(anti)proton scattering at high energy 
based on the modern formulation of Pomeron and Odderon in terms of Wilson lines and generalized 
TMDs (GTMDs). We compute the helicity-dependent elastic amplitudes $\phi_{1,2,3,4,5}$ in this model 
and study their energy dependence from the nonlinear small-$x$ evolution equations.
While both Pomeron and Odderon contribute to helicity-flip processes in general, in the forward 
limit $t=0$ only the double helicity-flip amplitude $\phi_2$, dominated by the spin-dependent 
Odderon, survives. This may affect the extraction of the $\rho$ parameter as well as the total 
cross section in the LHC energy domain and beyond.
\end{abstract}

\maketitle

\section{Introduction}
\label{Sec:Intro}

The elastic proton-(anti)proton scattering at high energies becomes an important source of 
information about the multi-layer proton structure \cite{Dremin:2019swd}. While the gluon-driven exchanges are dominant at asymptotically high energies (small-$x$), an elastic scattering implies, 
at least, a (colour-singlet) pair of correlated gluons propagating in the $t$-channel known 
as the QCD Pomeron (see e.g.~Ref.~\cite{Forshaw:1997dc} and references therein), in analogy to 
the leading pole exchange with Regge trajectory of the highest intercept \cite{Donnachie:1985iz} 
(for more detailed on the Regge theory, see Ref.~\cite{Collins:1977jy}). Even larger numbers 
of interacting gluons can be exchanged in an elastic scattering process, but the role of such 
multi-gluon interactions in elastic scattering yet remains uncertain, particularly, from the 
QCD point of view.

An odd-number gluon exchange starting from the leading triple-gluon one corresponds to 
the crossing-odd Odderon contribution in the Regge picture \cite{Bartels:1980pe,Kwiecinski:1980wb} 
(see also Ref.~\cite{Bartels:1999yt}). It was proposed back in the 70s in Ref.~\cite{Lukaszuk:1973nt} 
that the Odderon contribution may be non-negligible compared to that of the Pomeron in 
the high-energy limit. However, while an experimental observation of the Odderon is yet unavailable,
an exact magnitude and characteristics of such an elusive effect from theoretical viewpoint 
remain largely unknown and are the subjects of an intense debate and even controversial 
statements in the literature. 

The recent outbreak of Odderon activity (see e.g.~Refs.~\cite{Khoze:2018bus,Martynov:2018nyb,Shabelski:2018jfq}) is largely triggered by the 
precision TOTEM data at the highest energy of the LHC, $\sqrt{s}=$ 13 TeV, on total 
$\sigma_{\rm tot}$ \cite{Antchev:2017dia} and differential $d\sigma/dt$ \cite{Antchev:2018edk} 
$pp$ cross sections, as well as on the real-to-imaginary ratio of the elastic nuclear amplitude 
at the optical point, the so-called $\rho$-parameter \cite{Antchev:2017yns}. Introducing 
the total helicity non-flip elastic amplitude $T(s,t)$ as a function of the total c.m. energy squared $s$ and four-momentum 
transfer squared $t$, the basic measurable quantities of the elastic scattering read
 \beq
 \frac{d\sigma}{dt} = \frac{(1+\rho(s,t)^2)}{16\pi s(s-4M^2)} ({\rm Im}T(s,t))^2 \,, 
 \qquad  \rho(s,t) =\frac{{\rm Re}T(s,t)}{{\rm Im}T(s,t)} \,,
 \label{sigma}
 \eeq
so that the $\rho$-parameter is related to the total and differential (at vanishing momentum 
transfer) cross sections as follows
  \beq
\left.\frac{d\sigma}{dt}\right|_{t=0} = \frac{1+\rho^2}{16\pi}\sigma_{\rm tot}^2 \,, \qquad 
\sigma_{\rm tot} = \frac{ {\rm Im}T(s,t=0)}{\sqrt{s}\sqrt{s-4M^2}} \,, \qquad \rho \equiv 
\rho(s,t=0) \,,
\label{t0}
\eeq
where $M$ is the proton mass.
The $\rho$-parameter is small at TeV energies, $\rho \sim 0.1$, and has been extracted by 
the TOTEM collaboration in Ref.~\cite{Antchev:2017yns} from the experimental 
data on $d\sigma/dt$ near $t\approx 0$ using the Coulomb-Nuclear Interference (CNI). As long 
as $\rho(s)$ is known with sufficiently high precision, Eq.~(\ref{t0}) is used to determine 
$\sigma_{\rm tot}(s)$. The dominating claim is that a growth of the total cross section with 
energy, together with a decreasing $\rho$-parameter, as well as a qualitative 
difference of differential cross sections of $pp$ and $p\bar p$ collisions \cite{Ster:2015esa,Antchev:2018rec,Csorgo:2018uyp}, all are associated with the Odderon effect. 
There are some concerns in the literature, however, 
about the validity of the experimental procedure of $\rho$ extraction (see e.g.~\cite{Petrov:2018xma,Pancheri:2018yhd}) and to the Odderon interpretation of its decrease 
with energy (see e.g.~Ref.~\cite{Shabelski:2018jfq,Gotsman:2018buo}), and hence more care 
is needed to justify the magnitude and the significance of the Odderon effect in the $\rho$
measurement. In off-forward kinematics, a substantiated claim about the Odderon effect and 
its significance is made recently from the shape analysis of the elastic differential $pp$ 
and $p\bar p$ cross sections based upon their scaling properties in Ref.~\cite{Csorgo:2019ewn}. 
In the current study, we instead consider a possible Odderon effect and its energy dependence 
at the optical point of vanishing $t\approx 0$ only and leave the analysis of $t$-dependence 
for a future work.

The usual rationale, similarly to the Pomeron, is that the Odderon is assumed to not flip 
the helicities of the scattered hadrons. Indeed, the existing theoretical formulations and 
the procedure of $\rho$ extraction from the experimental data itself heavily rely on the 
presumption about an absence or a large suppression of helicity-flip processes at high energies. 
In this work, we question this convention and, in particular, explore a viable possibility that 
the helicity-flip elastic amplitude may be non-negligible at high energies. To our knowledge, this 
has neither been confirmed nor disproved by direct experimental measurements in the TeV region. 
On the other hand, it has been suggested in the literature that the Odderon can contribute to
helicity-flip amplitudes \cite{Ryskin:1987ya,Buttimore:1998rj,Leader:1999ua}. Yet, the exact 
treatment of the problem has been difficult due to the lack of a systematic way to connect 
the Pomeron and Odderon with the spin degrees of freedom of the scattering (composite) 
particles such as protons in QCD\footnote{For an earlier discussion of the Pomeron helicity flip observables for phenomenological scalar, vector and tensor Pomeron-proton couplings, see Ref.~\cite{Ewerz:2016onn}}.

Recently, however, there has been a significant progress in our understanding of the interplay 
between the Odderon and the proton spin  \cite{Zhou:2013gsa,Szymanowski:2016mbq,Boer:2015pni,Dong:2018wsp,Yao:2018vcg,Boussarie:2019vmk}. 
In the Deep Inelastic Scattering (DIS) at small-$x$, the Color Glass Condensate (CGC) framework \cite{Gelis:2010nm} provides a consistent description of the Pomeron and Odderon in terms of 
Wilson line correlators. Their couplings with various proton polarization states can be 
completely parametrized by the generalized transverse momentum dependent distributions 
(GTMDs) \cite{Boussarie:2019vmk}. Indeed, the gluon Sivers function \cite{Sivers:1989cc} 
at small-$x$ is connected to the Odderon in the forward limit \cite{Zhou:2013gsa,Boer:2015pni,
Szymanowski:2016mbq} and participates in 
the proton helicity-flip reactions including the unpolarised elastic scattering 
processes. In particular, it has been observed that the so-called
spin-dependent Odderon \cite{Zhou:2013gsa} can flip the proton helicity even in 
the forward limit, and this effect can survive at high energies since the Odderon 
intercept is exactly equal to unity \cite{Bartels:1999yt}.

Motivated by these developments, in this paper we introduce a new model of near-forward 
elastic proton-proton scattering designed for the TeV region and beyond. By treating one of 
the protons within the quark-diquark model, we can devise a setup analogous to DIS in the 
so-called dipole frame. In this frame, helicity-flip amplitudes can be calculated by exchanging 
the spin-dependent Pomeron and Odderon. We then study their energy dependence at $t=0$ by numerically solving the small-$x$ evolution equations for Pomeron and Odderon. Of course, in near-forward $pp$ scattering there is 
no apparent hard scale (like the photon virtuality $Q^2$ in DIS) which guarantees the use 
of perturbative approaches. However, in the TeV region one can  consider the saturation momentum $Q_s$ as 
a dynamically generated hard scale.

The paper is organised as follows. In Sect.~\ref{Sec:helicity-amps}, we introduce the basic helicity amplitudes of the elastic $pp$ scattering and discuss their role at high energies. In Sect.~\ref{Sec:q-qq_model}, we derive the helicity amplitudes in the quark-diquark dipole model and discuss their main properties. In Sect.~\ref{Sec:energy-dep}, energy dependence of the helicity amplitudes and their ratios is numerically studied from the nonlinear small-$x$ evolution equations. Finally, a brief summary and concluding remarks are given in Sect.~\ref{Sec:conclusions}.

\section{Helicity amplitudes}
\label{Sec:helicity-amps}

\begin{figure}[t]
\begin{center}
\includegraphics[width=7cm]{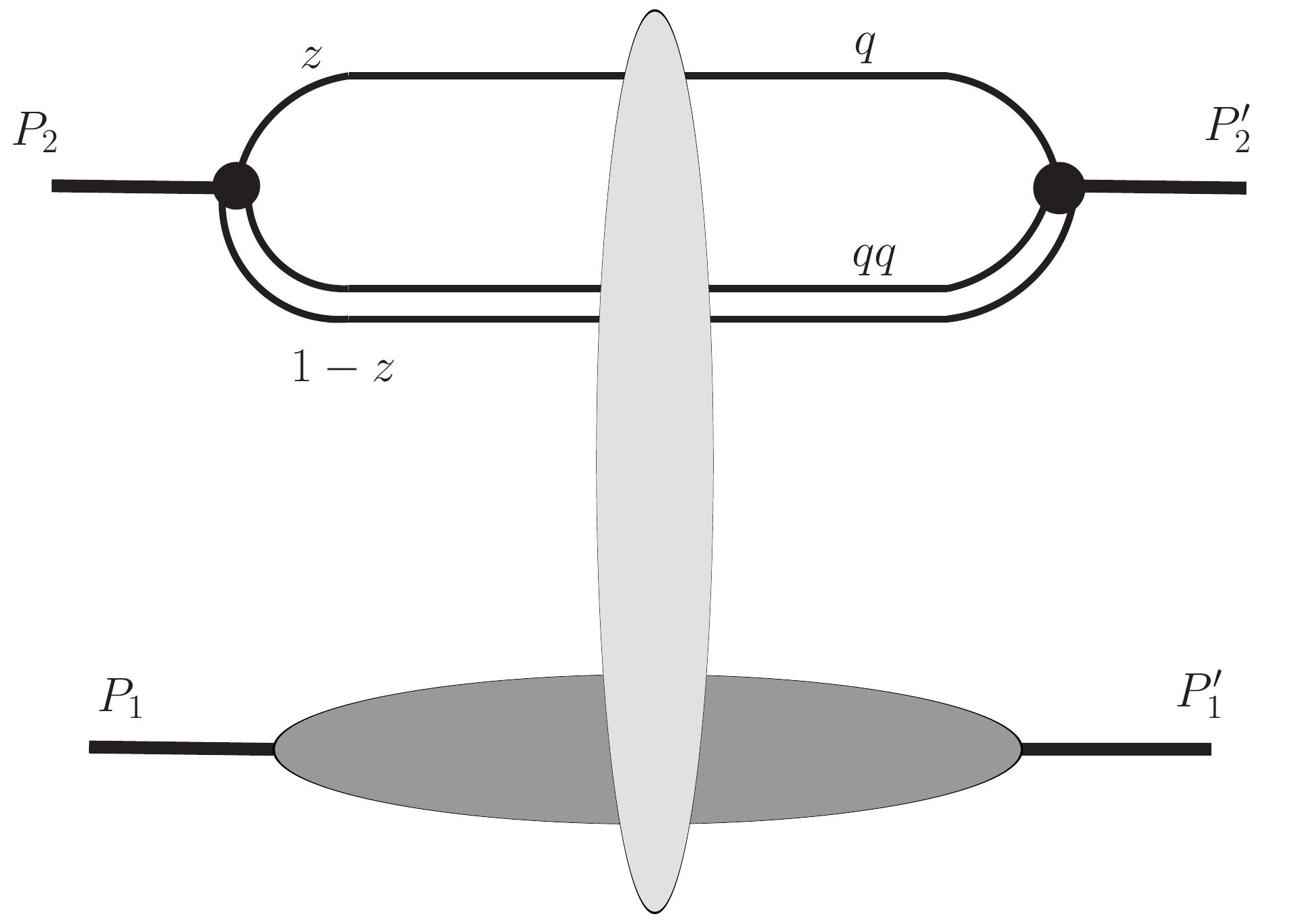}
\caption{Schematic illustration of the light-front dipole picture 
of elastic $pp$ scattering driven by a quark-diquark dipole scattering
off a proton target by means of the Pomeron and Odderon exchanges 
in the $t$-channel commonly denoted by a vertical grey blob.} 
\label{elastic-pp}
\end{center}
\end{figure}

Consider near-forward proton-proton elastic scattering $P_1P_2\to P'_1P'_2$ at high energies schematically shown in Fig.~\ref{elastic-pp}, with 4-momenta satisfying 
\begin{eqnarray}
P_1^\mu \approx \delta^\mu_+P_1^+ \,, \qquad P_2^\mu \approx \delta^\mu_-P_2^- \,, \qquad \Delta^\mu=P'^\mu_1-P_1^\mu = P_2^\mu-P_2'^\mu \approx \delta^\mu_i\Delta^i_\perp \,,
\end{eqnarray}
where $i=1,2$ denotes the transverse momentum components. 
We introduce the spin-dependent elastic amplitudes $\langle \lambda'_1\lambda'_2|T|\lambda_1\lambda_2\rangle$ \cite{Jacob:1959at,Goldberger:1960md,Leader:2001gr} where $\lambda=2h=\pm 1$ represents the helicity of each proton (multiplied by two, for convenience). These helicity amplitudes depend on  $s\approx 2P_1^+P_2^-$, $t\approx -\Delta^2_\perp$ as well as the azimuthal angle $\varphi={\rm Arg}(\Delta_\perp^1+i\Delta_\perp^2)$. We factor out the $\varphi$-dependence as\footnote{The exact phase factor depends on one's convention when defining the nucleon spinors, and the one we adopt here  may differ from those in the literature. Of course, the overall phase is unobservable and physically unimportant.}
\beq
\langle \lambda'_1\lambda'_2|T|\lambda_1\lambda_2\rangle \equiv e^{\frac{i}{2}(\lambda_1-\lambda_2-\lambda'_1+\lambda'_2)\varphi}\langle \lambda'_1\lambda'_2|\tilde{T}|\lambda_1\lambda_2\rangle, \label{factor}
\eeq
and switch to the commonly used notation 
\beq
&&8\pi \phi_1(s,t)=\langle ++|\tilde T|++\rangle, \quad 8\pi \phi_2(s,t) = \langle ++|\tilde T|--\rangle, \quad 8\pi\phi_3(s,t)=\langle +-|\tilde T|+-\rangle, 
\nonumber \\ 
&&8\pi \phi_4(s,t)=\langle +-|\tilde T|-+\rangle, \quad 8\pi\phi_5(s,t)=\langle ++|\tilde T|+-\rangle.
\eeq 
$\phi_{1,3}$ are the helicity non-flip amplitudes, $\phi_{2,4}$ are the double helicity-flip amplitudes and $\phi_5$ is the single helicity-flip amplitude. They are normalized such that the elastic differential cross section reads
\beq
\frac{d\sigma}{dt} = \frac{2\pi}{s(s-4M^2)}\left(|\phi_1|^2 
+ |\phi_2|^2 + |\phi_3|^2 + |\phi_4|^2 + 4|\phi_5|^2\right) 
\,, \label{dt}
\eeq
while the total cross section is 
\beq
\sigma_{\rm tot}=\frac{4\pi}{\sqrt{s}\sqrt{s-4M^2}}{\rm Im} (\phi_1(s,0)+\phi_3(s,0)) \,. \label{tot}
\eeq

A general argument shows that $\phi_4\propto t$, $\phi_5\propto \sqrt{-t}$ as $t\to 0$, whereas $\phi_{1,2,3}$ go to a constant in this limit \cite{Wang:1966zza,Leader:2001gr}. (Here we focus on the QCD part of the amplitude. The QED part behaves differently, see Appendix A.) Given these limiting behaviors, it is convenient to rescale the spin-dependent amplitudes as \cite{Buttimore:1998rj}
 \beq
  \qquad r_2(s,t)= \frac{ \phi_2(s,t)}{2{\rm Im}\phi_+(s,t)} &=& R_2 + iI_2, \qquad r_4(s,t)=\frac{M^2\phi_4(s,t)}{-t\, {\rm Im} \phi_+(s,t)}= R_4+iI_4\,, \label{r2} \\ 
r_5(s,t)&=&\frac{M\phi_5(s,t)}{\sqrt{-t}{\rm Im} \phi_+(s,t)}= R_5+iI_5 \,, \nonumber
 \eeq
where
\beq
\phi_+(s,t) \equiv \frac{\phi_1(s,t)+\phi_3(s,t)}{2} \,.
\eeq
The complex functions $r_{i=2,4,5}$ have a finite limit as $t\to 0$. They can be experimentally accessed by measuring various spin asymmetries \cite{Buttimore:1998rj}. For example, $r_5$ is closely related to single spin asymmetry $A_N$, and $r_2$ is related to double spin asymmetry $A_{NN}$. The results from fixed-target experiments at RHIC at $\sqrt{s}=13.76$ GeV and 21.92 GeV  \cite{Alekseev:2009zza,Poblaguev:2019vho} indicate that the parameters $r_2,r_5$ are small, of order $10^{-3}$ in this low-energy region. There are also RHIC data in the collider mode at $\sqrt{s}=200$ GeV \cite{Adamczyk:2012kn,Adamczyk:2013pna}. The analysis mostly focused on $A_N$ and a rather small value of $r_5$ has been reported. 

At higher energies, however, nothing is known about the fate of the helicity-flip amplitudes since there is no polarized proton collider beyond the RHIC energies. They are rarely discussed in connection with the ongoing measurements at the LHC, or with the earlier measurements at the Tevatron. It is usually assumed, often without even mentioning it, that $\phi_1\approx \phi_3$ and $\phi_{2,4,5}\approx 0$ for {\it all} values of $t$. There is then only one (complex) amplitude $T=8\pi\phi_1$, and (\ref{dt}) and (\ref{tot}) reduce to the  formulas mentioned in the introduction. Yet, even in unpolarized scattering, the helicity-flip amplitudes affect the observables. In the presence of nonvanishing $\phi_2$, Eq.~(\ref{t0}) should be modified as
\beq
\left.\frac{d\sigma}{dt}\right|_{t=0} = \frac{\sigma_{\rm tot}^2}{16\pi}(1+\rho^2+2|r_2|^2) \,. 
\label{corr}
\eeq  
Also, in the non-forward scattering with $|t|>0$, $\phi_{2,4,5}$ amplitudes could affect the shape of $d\sigma/dt$, especially, in the dip region where $|\phi_{1,3}|$ become small.   

In the next section, we compute all the $\phi$'s in a model which incorporates Pomeron and Odderon in the Wilson line formulation of small-$x$ QCD. We do not make the usual assumption that the helicity-flip amplitudes $\phi_{2,4,5}$ are negligibly small. As we shall see very clearly below, the spin-dependent Pomeron and Odderon exchanges naturally generate non-negligible helicity-flip amplitudes. The latter are {\it a priori} not suppressed at high energies since they share the same energy dependence (`Regge intercept') as for the helicity-conserving ones.

\section{Elastic scattering in the quark-diquark model}
\label{Sec:q-qq_model}

In this section, we calculate the helicity amplitudes $\phi_{1,\dots,5}$ in the dipole model of high-energy $pp$ (and $p\bar{p}$) scattering illustrated in Fig.~\ref{elastic-pp}. Our setup is similar to the description of DIS at small-$x$ in the so-called `dipole frame' where the virtual photon fluctuates into a quark-antiquark pair long before interacting with the target proton. Specifically,  we work in an asymmetric frame in which $P_1^+ \gg P_2^-$. The `slow', left-moving proton 2 is treated in the quark-diquark model \cite{Brodsky:2000ii}. It fluctuates into a quark and a scalar diquark, and the pair interacts with 
the shockwave created by the `fast' proton 1 in the eikonal approximation. The corresponding scattering amplitude is given by 
\begin{eqnarray}
T(s,t)=2 is\int \frac{d^2r_\perp}{4\pi} 
 \int_0^1 \frac{dz}{z(1-z)}
\Psi^*(r_\perp,z,\lambda_2')N(r_\perp,\Delta_\perp,\lambda_1,\lambda_1') \Psi(r_\perp,z,\lambda_2) \,.
\end{eqnarray}
Here, $\Psi$ is the light-front wave function of the proton 2 fluctuation into a $q-qq$ pair to be specified shortly, $\lambda_{1,2}$ and $\lambda'_{1,2}$ denote helicities of protons 1,2 in the initial and final states, respectively, $r_\perp$ is the transverse distance between the quark and the diquark, $z$ is the longitudinal momentum fraction of the proton 2 carried by the quark, and $N$ is the so-called dipole scattering amplitude defined by
\begin{eqnarray}
2P^+ 2\pi \delta(P^+-P^{'+}) N(r_\perp,\Delta_\perp,\lambda_1,\lambda'_1)\equiv \langle P'_1, \lambda'_1 |1-\frac{1}{N_c}
{\rm Tr} U(r_\perp/2) U^\dagger(-r_\perp/2) | P_1,\lambda_1 \rangle , \label{dipole}
\end{eqnarray}
in terms of a lightlike Wilson line in the fundamental representation 
\beq
U(x_\perp)={\rm P}\exp\left(ig\int dz^-A^+(z^-,x_\perp)\right),
\eeq
which describes the quark scattering off the target color field, and that for the diquark, $U^\dagger$, which has the same color representation as an antiquark. As usual, $g$ and $N_c=3$ denote the QCD coupling and number of colors. 

Following \cite{Boussarie:2019vmk}, we parametrize the dipole amplitude as 
\begin{eqnarray}
&& \!\!\!\!\!\!\!\!
\int d^2 r_\perp
e^{-ik_\perp \cdot r_\perp} N(r_\perp,\Delta_\perp,\lambda_1,\lambda_1')
\nonumber \\ &=&(2\pi)^4\delta^{(2)}(\Delta_\perp)\delta^{(2)}(k_\perp) \delta_{\lambda_1 ,\lambda_1'} - \frac{g^2(2\pi)^3}{8N_c M (k_\perp^2-\Delta_\perp^2/4)}
\bar u(P_1',\lambda_1') \left \{
\left [ f_{1,1}+i\frac{k_\perp \cdot \Delta_\perp}{M^2} g_{1,1}
\right ] \right .\
\nonumber \\ &&
\left .\ +i\frac{\sigma^{i+}}{ P_1^+} k_\perp^i\left[  \frac{k_\perp \cdot \Delta_\perp}{M^2}f_{1,2}
+ig_{1,2} \right]  +i\frac{\sigma^{i+}}{P_1^+}\Delta_\perp^i\left[f_{1,3}
+i\frac{k_\perp \cdot \Delta_\perp}{M^2}g_{1,3} \right ] \right \} u(P_1,\lambda_1)\,,  \label{ret}
\end{eqnarray}
where $f_{1,n}$ and $g_{1,n}$ ($n=1,2,3$) are functions of $k_\perp^2$, $\Delta_\perp^2$ and $|k_\perp \cdot \Delta_\perp|$ as well as the Bjorken-$x$ variable.  At small-$x$, they come from the real and imaginary parts of the operator ${\rm Tr}UU^\dagger$ and represent the Pomeron \cite{Balitsky:1995ub} and Odderon \cite{Hatta:2005as} exchanges, respectively. We note that the apparent pole at $k_\perp^2=\Delta_\perp^2/4$ is innocuous because $f$ and $g$ are proportional to $k_\perp^2-\Delta_\perp^2/4$, see for example, (\ref{gpd}) below.  

Let us now work out the product of spinors explicitly. Up to corrections of order $M/P^+_1$ and $\Delta_\perp/P^+_1$, we get 
 \begin{eqnarray}
&&\!\!\!\!\!\!\!\!\!\!\!\!
\bar u \left \{ f_{1,1}+i\frac{k_\perp \cdot \Delta_\perp}{M^2} g_{1,1}
\!+\! i\frac{\sigma^{i+}}{ P_1^+}k_\perp^i\left[  f_{1,2}\frac{k_\perp \cdot \Delta_\perp}{M^2}
+ ig_{1,2}  \right]\!+\!i\frac{\sigma^{i+}}{ P_1^+}\Delta_\perp^i \left[f_{1,3}
+i\frac{k_\perp \cdot \Delta_\perp}{M^2}g_{1,3} \right ]\right \} u
\nonumber \\
&\approx & 2M \delta_{\lambda_1 ,\lambda_1'}
\left [f_{1,1}(k_\perp)+i\frac{k_\perp \cdot \Delta_\perp}{M^2} g_{1,1}(k_\perp)\right ]
\nonumber \\&&
+2\lambda_1 \delta_{\lambda_1 ,-\lambda_1'}
k_\perp \cdot \epsilon_{\lambda_1}
\left [ \frac{k_\perp \cdot \Delta_\perp}{M^2} f_{1,2}(k_\perp)+ig_{1,2}(k_\perp) \right ]
\nonumber \\&& 
+2\lambda_1 \delta_{\lambda_1 ,-\lambda_1'}\Delta_\perp \cdot \epsilon_{\lambda_1} 
\left [ f_{1,3}(k_\perp)-\frac{1}{2}f_{1,1}(k_\perp)+i\frac{k_\perp \cdot \Delta_\perp}{M^2}\left(g_{1,3}(k_\perp)-\frac{1}{2}g_{1,1}(k_\perp)\right) \right ]\,,
\end{eqnarray} 
where we introduced the `polarization vector'
\beq
\epsilon_{\lambda} = (1,i\lambda), \qquad \Delta_\perp \cdot \epsilon_\lambda = \Delta_\perp^1+i\lambda \Delta^2_\perp = \sqrt{-t}e^{i\lambda \varphi}\,. 
\eeq
In the $r_\perp$-space, the parametrization takes the form,
\begin{eqnarray}
&&
N(r_\perp,\Delta_\perp,\lambda_1,\lambda_1')
 =(2\pi)^2\delta^{(2)}(\Delta_\perp)\delta_{\lambda_1 ,\lambda_1'} - \frac{g^2(2\pi)^3}{4N_cM}
\Biggl\{
M \delta_{\lambda_1 ,\lambda_1'}\left [\tilde f_{1,1}(r_\perp)
+\frac{\Delta_\perp \!\! \cdot r_\perp}{M^2r_\perp^2} \tilde g_{1,1}(r_\perp) \right ] \nonumber \\ &&  \quad  +
\lambda_1\delta_{\lambda_1 ,-\lambda_1'}\frac{ r_\perp \cdot \epsilon_{\lambda_1} }{r_\perp^2}
\left [ r_\perp\! \cdot \Delta_\perp \tilde f_{1,2}^b(r_\perp)+ \tilde g_{1,2}(r_\perp)\right ]
 \\&& 
 \quad +
\lambda_1\delta_{\lambda_1 ,-\lambda_1'}\Delta_\perp \cdot \epsilon_{\lambda_1} 
\left [\tilde f_{1,2}^a(r_\perp)+
 \tilde f_{1,3}(r_\perp)-\frac{\tilde f_{1,1}(r_\perp)}{2}+\frac{\Delta_\perp \!\! \cdot r_\perp}{M^2r_\perp^2} \left(\tilde g_{1,3}(r_\perp) -\frac{\tilde g_{1,1}(r_\perp)}{2}\right) 
 \right ]
\Biggr\}\,, \nonumber 
\end{eqnarray}
with
\begin{eqnarray}
\tilde f_{1,1}(r_\perp)&=&\int \frac{d^2 k_\perp}{(2\pi)^2} e^{ik_\perp \cdot r_\perp}
\frac{ f_{1,1}(k_\perp)}{k_\perp^2-\Delta_\perp^2/4}\,,
\\
\tilde f_{1,3}(r_\perp)&=&\int \frac{d^2 k_\perp}{(2\pi)^2} e^{ik_\perp \cdot r_\perp}
\frac{ f_{1,3}(k_\perp)}{k_\perp^2-\Delta_\perp^2/4}\,,
\\
\tilde f_{1,2}^a(r_\perp)&=&\int \frac{d^2 k_\perp}{(2\pi)^2} e^{ik_\perp \cdot r_\perp}
\left [ \frac{k_\perp^2}{M^2} -\frac{(r_\perp \cdot k_\perp)^2}{r_\perp^2M^2}\right ]
\frac{ f_{1,2}(k_\perp)}{k_\perp^2-\Delta_\perp^2/4}\,,
\\
\tilde f_{1,2}^b(r_\perp)&=&\int \frac{d^2 k_\perp}{(2\pi)^2} e^{ik_\perp \cdot r_\perp}
\left [ 2\frac{(r_\perp \cdot k_\perp)^2}{r_\perp^2M^2}-\frac{k_\perp^2}{M^2} \right ]
\frac{ f_{1,2}(k_\perp)}{k_\perp^2-\Delta_\perp^2/4}\,,
\end{eqnarray}
and
\begin{eqnarray}
\tilde g_{1,n}(r_\perp)&=&i\int \frac{d^2 k_\perp}{(2\pi)^2} e^{ik_\perp \cdot r_\perp}
k_\perp \cdot r_\perp \frac{ g_{1,n}(k_\perp)}{k_\perp^2-\Delta_\perp^2/4} \,,
\end{eqnarray}
for $n=1,2,3$. All functions defined above are real-valued.

The wave function of proton 2 in the quark-diquark model is given by (see Appendix B for the relevant Feynman rules)
\begin{eqnarray}
\Psi(r_\perp,z,\lambda_2)=-c_s \int \frac{d^2 l_\perp}{(2\pi)^2} e^{-ir_\perp \cdot l_\perp} \frac{z(1-z)
\bar u(z,l_\perp,\lambda_q)u(P_2,\lambda_2)}{l_\perp^2+\tilde M^2} \,,
\label{wf}
\end{eqnarray}
where $c_s$ is a constant normalisation. The constituent quark has momentum fraction $z$, transverse momentum 
$l_\perp$, mass $m_q$ and helicity $\lambda_q$.  The scalar diquark  has momentum fraction $\bar{z}=1-z$ and mass  $m_s$. Due to a finite binding energy, $m_q+m_s\ge M$, and this ensures that $\tilde M^2\equiv \bar z m_q^2 + z m_s^2-z\bar z M^2\ge 0$. After carrying out the integration over $l_\bot$, one can rewrite the wave function in Eq.~(\ref{wf}) as,
\begin{eqnarray}
\Psi(r_\perp,z,\lambda_2)= \frac{-c_s \sqrt{z}\bar z}{2\pi \sqrt{2}} \left[
\delta_{\lambda_q,\lambda_2}\left(Mz+m_q\right) K_0(\tilde M |r_\perp|)
- \delta_{\lambda_q,-\lambda_2}\lambda_2 r_\perp \cdot \epsilon^*_{\lambda_2} 
\frac{i\tilde M}{|r_\perp|} K_1(\tilde M |r_\perp|) \right] \,.
\label{wfhxtilde}
\end{eqnarray}
This leads to the following expression for the wave function
squared in the forward limit,
\begin{eqnarray}
\sum\limits_{\lambda_q}\Psi(r_\perp,z,\lambda_2)\Psi^*(r_\perp,z,\lambda_2')
= c_s^2\frac{z \bar z^2}{(2\pi)^2} \left( \delta_{\lambda'_2,\lambda_2} \frac{\Phi_n(r_\perp)}{M} -2i\lambda_2 \delta_{\lambda'_2,-\lambda_2} 
r_\perp \cdot \epsilon^*_{\lambda_2} \Phi_f(r_\perp)\right) \,,
\label{dcstilde}
\end{eqnarray}
in terms of the helicity flip and helicity non-flip parts of the wave function 
\begin{eqnarray} \label{eq:Phi-n}
\Phi_{n}(r_\perp)&=& M
 \left [\left( Mz+m_q \right)^2K_0^2(\tilde M |r_\bot|) +\tilde M^2 K_1^2(\tilde M |r_\bot|) \right ] \,,
 \\
\Phi_{f}(r_\perp)&=& \left( Mz+m_q  \right)     K_0(\tilde M |r_\bot|)
  \frac{\tilde M}{|r_\bot|}K_1(\tilde M|r_\bot|) \,, \label{eq:Phi-f}
\end{eqnarray}
respectively. In the non-forward case, a nontrivial phase $e^{i \left(z-\frac{1}{2}\right)\Delta_\perp \cdot r_\perp}$ emerges\footnote{For the reader's convenience, here we briefly recapitulate the discussion in Ref.~\cite{Hatta:2017cte}. The non-forward amplitude in dipole models typically has the following structure in impact parameter space $b_\perp$,
\begin{equation} 
T(b_\perp) = \int d^2\Delta_\perp e^{-ib_\perp \cdot \Delta_\perp} T(\Delta_\perp)
\sim \int d^2r_\perp |\Psi(\Delta_\perp=0)|^2 N(b_\perp-zr_\perp) \,,
\end{equation}
where $N$ is the dipole scattering amplitude. The shift $b_\perp \to b_\perp - zr_\perp$ is caused by the phase factor $e^{izr_\perp \cdot \Delta_\perp}$ which generically appears in non-forward impact factors \cite{Bartels:2003yj}, see Eq.~(\ref{gpd}) below for example. This implies that $b_\perp+(1-z)r_\perp$ and $b_\perp-zr_\perp$ can be interpreted as the coordinate of the quark and antiquark (or diquark), respectively. We can thus identify
\begin{equation}
N(b_\perp-zr_\perp) =\left\langle 1-\frac{1}{N_c}{\rm Tr}\, U(b_\perp+(1-z)r_\perp)U^\dagger(b_\perp-zr_\perp)\right\rangle \,.
\end{equation}
This gives 
\beq
T(\Delta_\perp) &\sim& \int d^2b_\perp e^{ib_\perp\cdot \Delta_\perp}\int d^2r_\perp |\Psi|^2 \left\langle {\rm Tr}\, U(b_\perp+(1-z)r_\perp)U^\dagger(b_\perp-zr_\perp) \right\rangle\nn 
&=& \int d^2b_\perp e^{ib_\perp \cdot \Delta_\perp} \int d^2r_\perp e^{i \left(z-\frac{1}{2}\right)\Delta_\perp \cdot r_\perp} |\Psi|^2 \left\langle{\rm Tr}\, U(b_\perp+r_\perp/2)U^\dagger(b_\perp-r_\perp/2)\right\rangle \,.
\eeq
} \cite{Hatta:2017cte}.
We keep the subleading terms up to quadratic order in  $\Delta_\perp$, so in practice we use
\beq
\sum\limits_{\lambda_q}\Psi\Psi^* \to \sum\limits_{\lambda_q}\Psi\Psi^* e^{i \left(z-\frac{1}{2}\right)\Delta_\perp \cdot r_\perp} \approx \sum\limits_{\lambda_q}\Psi\Psi^* \left [1+i z_*\Delta_\perp \cdot r_\perp-\frac{z_*^2(\Delta_\perp \cdot r_\perp)^2}{2} \right ] \,,
\eeq 
where we used the abbreviation $z_*=z-1/2$. Assembling the above pieces together, we get
\begin{eqnarray}
&& \!\!\!\!\!\!
\int d^2 r_\perp
\Psi^*(r_\perp,z,\lambda_2')N(r_\perp,\Delta_\perp,\lambda_1,\lambda_1') \Psi(r_\perp,z,\lambda_2)
\nonumber \\ 
&& =
-\frac{\pi g^2c_s^2 }{2N_cM}z \bar z^2
\int d^2r_\perp 
\Biggl\{ \delta_{\lambda_2',\lambda_2}\delta_{\lambda_1 ,\lambda_1'}\Phi_{n}
 \left[ \tilde H \left(1- \frac{z_*^2}{4}\Delta_\perp^2 r_\perp^2 \right)-\delta^{(2)}(\Delta_\perp) \frac{2N_c}{g^2 \pi} +\frac{iz_*\Delta_\perp^2 }{2M^2}   \tilde g_{1,1} \right ]
 \nonumber \\
&&+ \delta_{\lambda_2',-\lambda_2}\delta_{\lambda_1 ,-\lambda_1'} \delta_{\lambda_1,\lambda_2}  \Phi_{f}
\left [-2i\tilde g_{1,2}  -i \frac{\Delta_\perp^2}{2M^2}\left ( 2\tilde g_{1,3}-\tilde g_{1,1}-z_*^2 M^2r_\perp^2 \tilde g_{1,2} \right ) + \frac{z_*\Delta_\perp^2 r_\perp^2 }{2}\tilde E \right ] \nonumber \\
&& -  \delta_{\lambda_2',-\lambda_2}\delta_{\lambda_1 ,-\lambda_1'} \delta_{\lambda_1,-\lambda_2}  \Phi_{f}
  (\Delta_\perp \cdot \epsilon_{\lambda_1} )^2
 \left [ \frac{z_*r_\perp^2}{2}\tilde{E}-\frac{i}{2M^2}\left(2\tilde g_{1,3}-\tilde g_{1,1}-\frac{z_*^2M^2 r_\perp^2}{2}\tilde g_{1,2}\right) \right ]
\nonumber \\
&&+ 
\delta_{\lambda_2',\lambda_2}\delta_{\lambda_1 ,-\lambda_1'}
\lambda_1\Delta_\perp \cdot \epsilon_{\lambda_1}   \frac{\Phi_{n}}{2M}
\left [\tilde{E} +iz_*\tilde g_{1,2}
 \right ]
\nonumber \\
&&+ \delta_{\lambda_2',-\lambda_2}\delta_{\lambda_1 ,\lambda_1'}
\lambda_2\Delta_\perp \cdot \epsilon^*_{\lambda_2}   \Phi_{f}
\left [z_*Mr_\perp^2 \tilde H -\frac{i\tilde g_{1,1}}{M}\right ]\Biggr\} \,, \label{as}
\end{eqnarray}
where we defined 
\beq
 \tilde H \equiv \tilde{f}_{1,1} = \int \frac{d^2k_\perp}{(2\pi)^2} e^{ik_\perp \cdot r_\perp}  \frac{f_{1,1}(k_\perp,\Delta_\perp)}{k_\perp^2-\Delta^2_\perp/4} \,,
\eeq
and
\beq
\tilde{E} \equiv  2\tilde f_{1,3}- \tilde f_{1,1}
+2\tilde f_{1,2}^a+\tilde f_{1,2}^b &=& \int \frac{d^2k_\perp}{(2\pi)^2} e^{ik_\perp \cdot r_\perp} \frac{-f_{1,1}(k_\perp)+2f_{1,3}(k_\perp)  + \frac{k_\perp^2}{M^2}f_{1,2}(k_\perp)}{k_\perp^2 -\Delta_\perp^2/4} \nonumber \\ 
&\equiv & \int \frac{d^2k_\perp}{(2\pi)^2} e^{ik_\perp \cdot r_\perp} \frac{E(k_\perp,\Delta_\perp)}{k_\perp^2 -\Delta_\perp^2/4} \,.
\eeq
Above, $f_{1,1}$ and $E$ are nothing but the GTMD version of the GPDs of the fast proton (c.f., Eq.~(4.48) of Ref.~\cite{Meissner:2009ww}) normalized as 
\beq
\int d^2k_\perp f_{1,1}(k_\perp,\Delta_\perp) =  H(t) \,, \qquad 
\int d^2k_\perp E(k_\perp,\Delta_\perp) = E(t) \,,
\eeq
where $H$ and $E$ are the standard gluon GPDs. Note that, since we are colliding identical particles, by symmetry the coefficients of $\delta_{\lambda_2',-\lambda_2}\delta_{\lambda_1 ,\lambda_1'}$ and $\delta_{\lambda_2',\lambda_2}\delta_{\lambda_1 ,-\lambda_1'}$ have to be equal (up to a sign and trivial relabeling). However, in the asymmetric frame in which we are working, this is not obvious at first sight. While we do not have an explicit proof, we nevertheless argue that the two expressions are indeed equivalent. The functions $\Phi_n$ and $\Phi_f$ introduced in Eqs.~(\ref{eq:Phi-n}) and (\ref{eq:Phi-f}) are related to the helicity non-flip and helicity flip parts of the gluon GTMD of the slow proton, respectively,
\beq
G(k_\perp, \Delta_\perp) \propto C_F \alpha_s^2 \int \frac{dz}{z\bar{z}}\int d^2r_\perp \sum\Psi\Psi^* e^{iz\Delta_\perp \cdot r_\perp} \frac{k_\perp^2-\Delta_\perp^2/4}{(\Delta_\perp/2-k_\perp)^2(\Delta_\perp/2+k_\perp)^2} \nonumber \\  \times (1-e^{-ir_\perp\cdot (\Delta_\perp/2-k_\perp)})(1-e^{-ir_\perp\cdot (\Delta_\perp/2+k_\perp)}) \,.
\label{gpd}
\eeq
The $t$-channel gluon propagators in Eq.~(\ref{gpd}) (as well as the small-$x$ evolution) are absorbed into $\tilde E$ and $\tilde H$. Thus, the terms proportional to $\Phi_n\tilde{E}$ and $\Phi_f \tilde{H}$ in the last two lines of Eq.~(\ref{as}) are both the convolution of the $H$-type GTMD of one proton and the $E$-type GTMD of the other proton, and are thus equal. A similar argument applies to the imaginary parts proportional to $\Phi_n \tilde g_{1,2}$ and $\Phi_f \tilde g_{1,1}$. Although there is in general no relation between $\tilde g_{1,2}$ and $\tilde g_{1,1}$, they satisfy the same evolution equation. The only difference is the way the $t$-channel Odderon amplitude ${\cal T}_O$ couples to the proton, and this coupling is proportional to $\Phi_f$ and $\Phi_n$, respectively, cf., Ref.~\cite{Szymanowski:2016mbq}. Thus, the imaginary terms in the last two lines of Eq.~(\ref{as}) both have the structure $\Phi_n\otimes {\cal T}_O \otimes \Phi_f$, and are thus equivalent. After removing the phase according to Eq.~(\ref{factor}), we arrive at 
\beq
\phi_1=\phi_3=-  \frac{isg^2c_s^2 }{8N_cM}\int_0^1 dz \bar{z}\int \frac{ d^2r_\perp}{4\pi} \Phi_n \left[\tilde H \left(1- \frac{z^2_*}{4}\Delta_\perp^2 r_\perp^2 \right)- \delta^{(2)}(\Delta_\perp) \frac{2N_c}{g^2 \pi} +\frac{iz_*\Delta_\perp^2 }{2M^2}   \tilde g_{1,1} \right]\,, \label{phi1} 
\eeq
\beq
\phi_2=-  \frac{sg^2c_s^2 }{4N_cM}\int_0^1 dz \bar{z} \int \frac{ d^2r_\perp}{4\pi}  \Phi_f \left[ \tilde{g}_{1,2}+\frac{\Delta_\perp^2}{4M^2}  (2\tilde g_{1,3}-\tilde g_{1,1}-z_*^2M^2r_\perp^2\tilde g_{1,2}) + i\frac{z^2_*}{4}\Delta_\perp^2 r_\perp^2 \tilde E \right]\,, \label{phi2}
\eeq
\beq
\phi_4= \frac{is g^2c_s^2}{16N_cM} (-t)\int dz \bar{z} \int \frac{d^2r_\perp}{4\pi} \Phi_{f}
\left (z_*  r_\perp^2 \tilde{E}-\frac{i}{M^2}\left(2\tilde g_{1,3}-\tilde g_{1,1}-\frac{z_*^2M^2r_\perp^2}{2}\tilde g_{1,2}\right) \right )\,,
\eeq
\beq
\phi_5= \frac{isg^2c_s^2 }{16N_cM}\sqrt{-t} \int_0^1 dz \bar{z}\int \frac{ d^2r_\perp}{4\pi} 
 z_*\left( 2\Phi_f Mr_\perp^2\tilde H  -\frac{i}{M}\Phi_n \tilde g_{1,2}\right) \,. \label{phi5model}
\eeq
The sign in front of $\tilde g_{1,2}$ in Eq.~(\ref{phi5model}) has been fixed using the relation $\langle ++|\tilde T|+-\rangle = -\langle ++|\tilde{T}|-+\rangle$ \cite{Buttimore:1978ry}\footnote{Since the last two lines of Eq.~(\ref{as}) are equivalent as we have argued, we may choose any linear combination of $\{\tilde H, \tilde E\}$ and $\{\tilde g_{1,1}, \tilde g_{1,2}\}$ in Eq.~(\ref{phi5model}). Here, we chose the set $\{\tilde H, \tilde g_{1,2}\}$ merely because we have numerical results available for these distributions.}.   
We immediately notice that $\phi_{1,3}$ are purely imaginary and $\phi_2$ is purely real. Therefore, the usual $\rho$-parameter (\ref{sigma}) vanishes at $t=0$ in this model. Away from $t=0$, the $\rho$-parameter is dominated by the spin-independent Odderon $\tilde g_{1,1}$. We also see that the Pomeron ($\tilde H,\tilde E$) and Odderon ($\tilde g_{1,2,3}$) contributions are always relatively imaginary. This means that there is no interference when squaring the amplitudes $|\phi_i|^2$, and $d\sigma/dt$ is insensitive to the sign of $\tilde g_{1,2,3}$. In other words, $d\sigma/dt$ is identical for $pp$ and $p\bar{p}$ scatterings in this model.

Recently, there are indications that the difference $d\sigma^{pp}/dt-d\sigma^{p\bar{p}}/dt$ is nonvanishing from an analysis of the LHC and Tevatron data \cite{Martynov:2017zjz,Csorgo:2019ewn}. In order to explain this,  the Odderon has to have a small imaginary part (and the Pomeron has a small real part). It may be possible to generalize our model to accommodate this effect, for example, by using the dispersion relation or invoking  Regge theory or the AdS/CFT correspondence \cite{Avsar:2009hc}. This is however beyond the scope of this work.

As for the ratios (\ref{r2}), we get 
 \beq
r_2(s,t=0) = \frac{ \int dz \bar{z} \int d^2r_\perp \Phi_f\tilde g_{1,2}}{\int dz \bar{z} \int d^2r_\perp \Phi_n \left(\tilde H - \frac{{\cal A}N_c}{2\pi^3g^2}\right)} = R_2 + iI_2 \,,
\label{eq:r2}
\eeq
\beq
r_4(s,t\approx 0) = \frac{ \int dz \bar{z} \int d^2r_\perp \Phi_f\left [-2\tilde g_{1,3}+\tilde g_{1,1}+\frac{z_*^2M^2r_\perp^2}{2}\tilde g_{1,2} -iz_* M^2 r_\perp^2 \tilde{E} \right ]}{2 \int dz \bar{z} \int d^2r_\perp \Phi_n \left(\tilde H - \frac{{\cal A}N_c}{2\pi^3g^2}\right)} = R_4 + iI_4 \,,
\eeq
\beq
r_5(s,t\approx 0) = \frac{ \int dz  \bar{z}\int d^2r_\perp  
z_* \left[ -\Phi_n \tilde g_{1,2} - 2i\Phi_f M^2r_\perp^2 \tilde H \right] }{ 2 \int dz \bar{z} \int d^2r_\perp \Phi_n \left(\tilde H - \frac{{\cal A}N_c}{2\pi^3g^2}\right)} = R_5 + iI_5 \,.
\label{eq:r5}
 \eeq
Since $\phi_{4,5}$ vanish at $t=0$, $r_{4,5}$ are not well-defined at $t=0$, and of course measurements are always performed at $t\neq 0$. On the other hand, $r_2$ has a well-defined limit $t\to 0$, and there we need to subtract $(2\pi)^2\delta^2(\Delta_\perp=0) \equiv {\cal A}$, the transverse area of the proton, from $\tilde H$ in the denominator. This converts the $S$-matrix $(\tilde H$) into the $T$-matrix, and is crucial for the $r_\perp$ integral to converge at small $r_\perp$. [Note that $\Phi_{n,f}(r_\perp) \sim 1/r_\perp^2$ at small-$r_\perp$.] When $t$ is nonzero, the subtraction is absent but there is no convergence problem since $\tilde H$ vanishes at $r_\perp=0$ if $\Delta_\perp\neq 0$. However, we can keep this subtraction in the denominator of $r_{4,5}$ and evaluate it at $t=0$ thanks  to the fact that the limit ${\rm Im}\, \phi_1(t\to 0)$ is smooth. Also, $\tilde{H}$ in the numerator of $r_5$ can be safely evaluated at $t=0$ since the factor $r_\perp^2$ kills the divergence at $r_\perp=0$. In the present work, these tricks are crucial for the numerical study in the next section since we do not have a numerical solution of $\tilde{H}$ at  $t\neq 0$.

We see that the real parts $R_{2,4,5}$  entirely come from the Odderon. In particular, $R_2$ at $t=0$ is nonvanishing due to  the spin-dependent Odderon $g_{1,2}$, and this can contribute to the differential and total cross section according to Eq.~(\ref{corr}). The imaginary parts $I_{4,5}$ come from the Pomeron and $I_2$ vanishes in this model. It is interesting to notice that $I_5$ is parametrically of order unity if the typical value is $r_\perp \sim 1/M$. However, at high energy the integrand is more localized at small-$r_\perp$, and then the factor $r_\perp^2$ leads to a suppression of $I_5$ (see below). We also expect $|I_5|\gg |I_4|$ assuming $|H|\gg |E|$.

\section{Energy dependence of the helicity amplitudes}
\label{Sec:energy-dep}

In this section, 
we study the center-of-mass energy $\sqrt{s}$ dependence of the helicity amplitudes $\phi_i$ and their ratios obtained in the previous section.  $f_{1,n}$ and $g_{1,n}$ are the real and imaginary parts of the dipole scattering amplitude (\ref{dipole}), respectively. The latter satisfies   the Balitsky-Kovchegov (BK) equation \cite{Balitsky:1995ub,Kovchegov:1999yj} which is an evolution equation in $\ln s$ including the gluon saturation effect. Thus, the Pomeron and Odderon amplitudes can be obtained from the real and imaginary parts of the BK equation with appropriate initial conditions \cite{Kovchegov:2003dm,Hatta:2005as}. We restrict ourselves to the forward limit $\Delta_\perp=0$, which means that we concentrate on $f_{1,1}$ and $g_{1,2}$. Solving the BK equation with finite $\Delta_\perp$ is numerically more involved, and to our knowledge this has not been done for the Odderon. 

Admittedly, the use of the BK equation for our problem must be  legitimately criticized.  Being an equation originally derived in perturbation theory, in principle the BK equation can only apply to processes which involve a hard scale. However, in near-forward elastic $pp$ scattering, apparently there is no such hard scale.  Yet, the idea of gluon saturation and the Color Glass Condensate \cite{Gelis:2010nm} is that at asymptotically high energies, the gluon distribution in the colliding particles is characterized by a dynamically generated hard scale, called the saturation momentum $Q_s(s)$ which is a rapidly increasing function of $\sqrt{s}$. There are indications that already in $pp$ collisions at the LHC, $Q_s$ is large enough so that the framework is applicable, see for example, \cite{McLerran:2010ex}. This partly justifies our approach at least for the Pomeron, and allows us to calculate the perturbative part of the growth of the total cross section with energy. Of course there are also nonperturbative contributions to the total cross section, but in our model these are absorbed into the parameter ${\cal A}$.   As a matter of fact, 
the same argument does not quite hold  for the Odderon. It has been noticed that the characteristic momentum scale of the Odderon amplitude does not grow like $Q_s$  \cite{Lappi:2016gqe,Yao:2018vcg}. Therefore,  the  results involving Odderon below are at best a crude estimate of the possible high energy behavior suggested by perturbation theory. In reality the dominance of the nonperturbative effects may be overwhelming.

We basically follow Ref.~\cite{Yao:2018vcg} for the numerical evaluation of $f_{1,1}$ and $g_{1,2}$, except that we now include the running coupling effect.  Ref.~\cite{Yao:2018vcg} considered a transversely polarized proton and studied the gluon Sivers function which is the forward limit of $g_{1,2}$. On the other hand, in our problem the proton is longitudinally polarized. We thus need a little spinor algebra to connect the two works.  
Let us return to Eq.~(\ref{ret}) and take the forward limit $\Delta_\perp=0$,
\beq
\int d^2r_\perp e^{-ik_\perp \cdot r_\perp }N(r_\perp) 
=(2\pi)^2\delta^{(2)}(k_\perp){\cal A} -\frac{g^2(2\pi)^3}{4N_cM k_\perp^2} \left(Mf_{1,1} +  \epsilon^{ij}k_\perp^i S^j_\perp    g_{1,2}\right) \,.
\eeq
Here we assume that the proton is transversely polarised, with the transverse spin vector $\vec S_\perp$ normalised as $|\vec S_\perp|=1$. In the $r_\perp$-space,
\beq
N(r_\perp)={\cal A} -\frac{g^2(2\pi)^3}{4N_cM} \left(M \tilde{H}(r_\perp) + i\frac{ \epsilon^{ij} S^i_\perp  r^j_\perp}{r_\perp^2} \tilde{g}_{1,2}(r_\perp)\right) \,.
\eeq
This can be written as (compare with Eq.~(5) of \cite{Yao:2018vcg})
\beq
S(\vec x_\perp,\vec y_\perp) = P(r_\perp) + i\vec S_\perp \times \vec r_\perp\, Q(r_\perp)\,, 
\qquad
\vec r_\perp \equiv \vec x_\perp - \vec y_\perp \,,
\eeq
where 
\beq
P(r_\perp)=\frac{g^2(2\pi)^3}{4N_c \cal A} \tilde{H}(r_\perp) \,, \qquad
Q(r_\perp) = \frac{g^2(2\pi)^3}{4N_cM{\cal A}} \frac{\tilde{g}_{1,2}(r_\perp)}{r_\perp^2} \,, \label{P-Q}
\eeq
for the Pomeron and the spin-dependent Odderon components of the dipole $S$-matrix,
respectively.
\begin{figure}[h]
\begin{center}
\includegraphics[width=11cm]{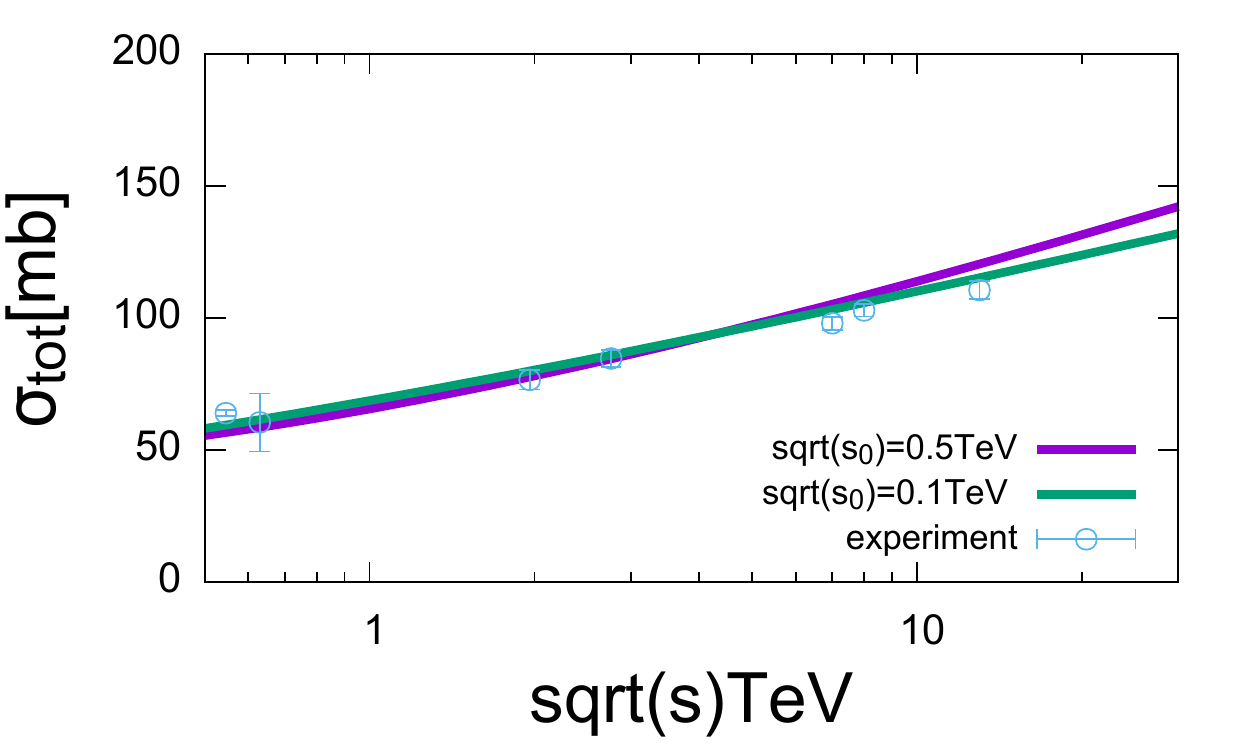}
\caption[]{The energy dependence of the total cross section.  
We take $Q_{s0} = 1.0$ GeV  
to fit  the experimental data. 
}
\label{figcrosssec}
\end{center}
\end{figure}

We compute $P(s,r_\perp)$ and 
$Q(s,r_\perp)$ as functions of the center-of-mass energy squared $s$ from the solution of the BK equation with running coupling as prescribed in Ref.~\cite{Balitsky:2006wa}. Then, using Eq.~(\ref{P-Q}) we 
access $\tilde{H}(r_\perp)$ and $\tilde{g}_{1,2}(r_\perp)$ that are further 
employed in computing $r_{2,5}$ through 
Eqs.~(\ref{eq:r2}) and (\ref{eq:r5}).  We adopt the following form for the coupling constant
\beq
	\alpha_s(r_\perp^2) = \frac{1}{b_0\log\left( \frac{4}{r_\perp^2 \Lambda} + a \right)},
	\eeq
with $b_0 = \frac{9}{4\pi}$ (corresponding to $n_f=3$), $\Lambda = 0.241~$GeV and $a = e^{\frac{8\pi}{9}}$.   
The initial conditions are given at the starting energy scale
$s_0$ as follows:
\beq
P(s_0,r_\perp) = e^{-r_\perp^2Q_{s0}^2/4}\,, \qquad 
Q(s_0,r_\perp) = \kappa Q_{s0}^3r_\perp^2 e^{-r_\perp^2Q_{s0}^2/4} \,.
\eeq
The initial saturation scale $Q_{s0}$ is expected to be around 1 GeV in the TeV region, while the strength of Odderon $\kappa$ is an unknown parameter including its sign (see, however, \cite{Zhou:2013gsa}) which should be  fitted to the data  \cite{Boussarie:2019vmk}. 
The other parameters in this model are $m_q$, $m_s$ and $c_s^2{\cal A}$ (only this product enters our  observables). We fix $m_q=0.3$ GeV and $m_s=M-m_q$, while $c_s^2{\cal A}$ is fitted to the total cross section.

\begin{figure}[h]
\begin{center}
\includegraphics[width=8cm]{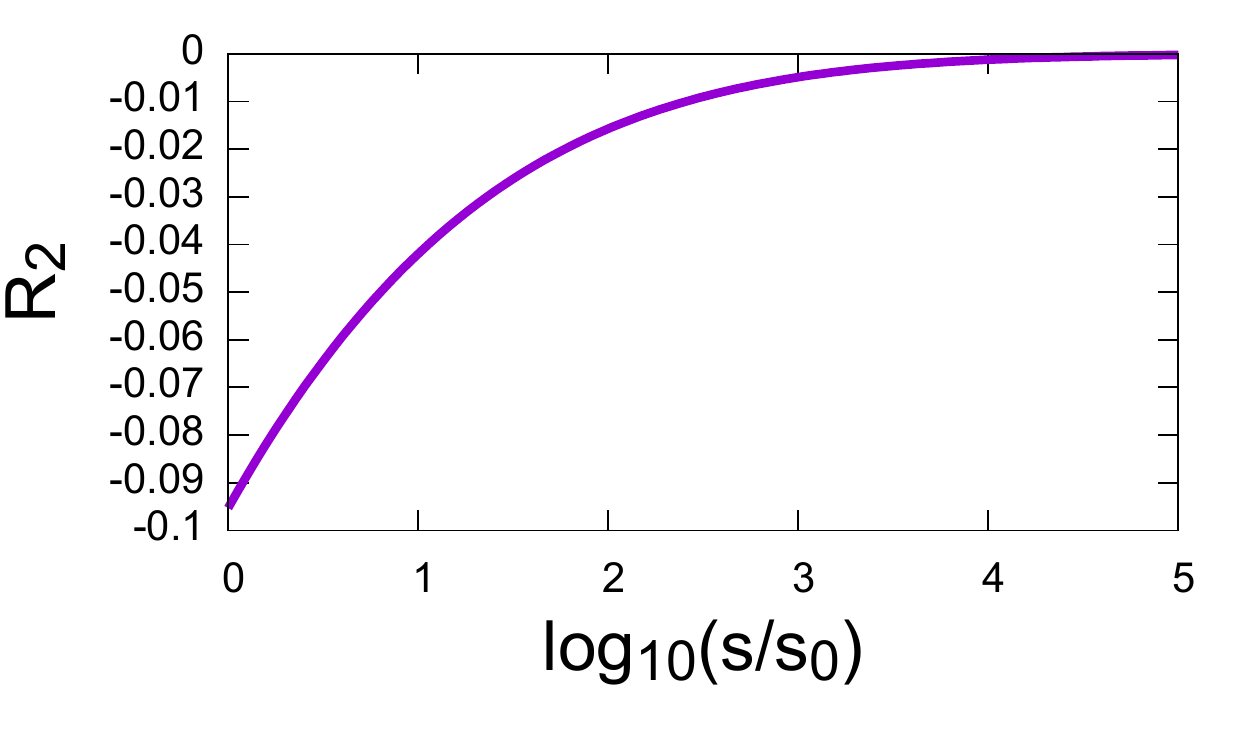}
\includegraphics[width=8cm]{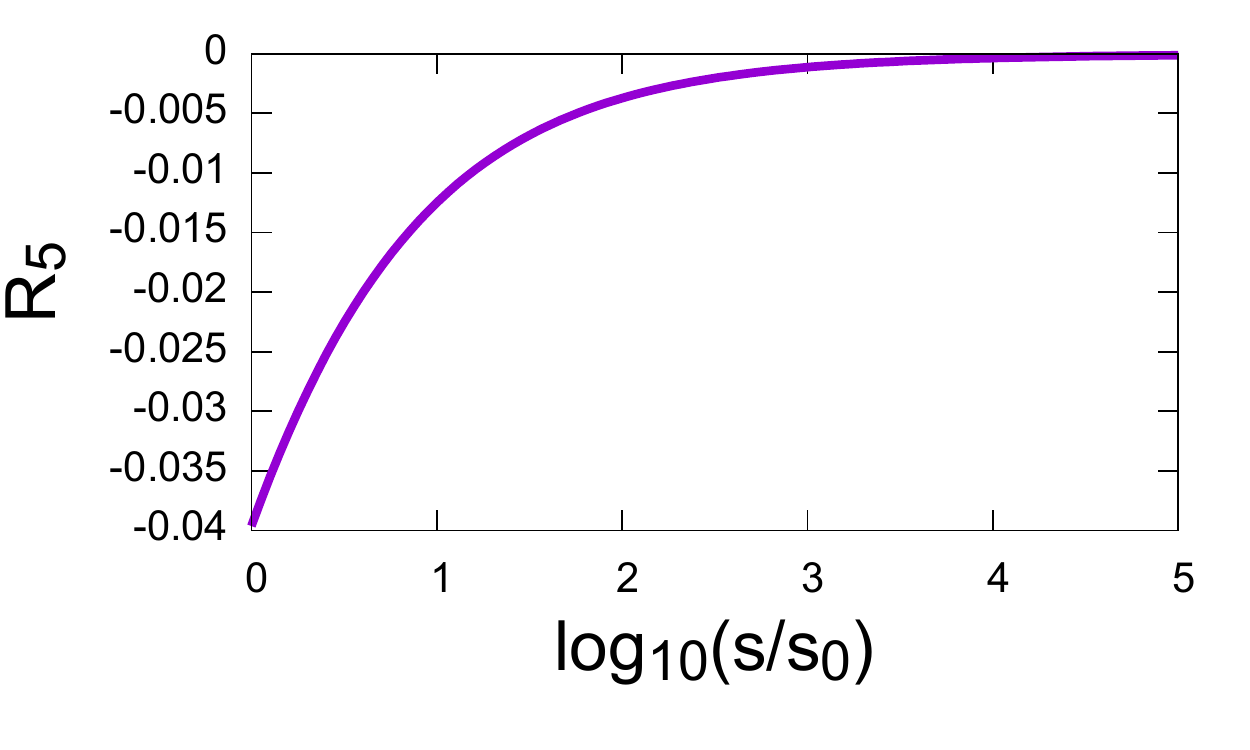}
\includegraphics[width=8cm]{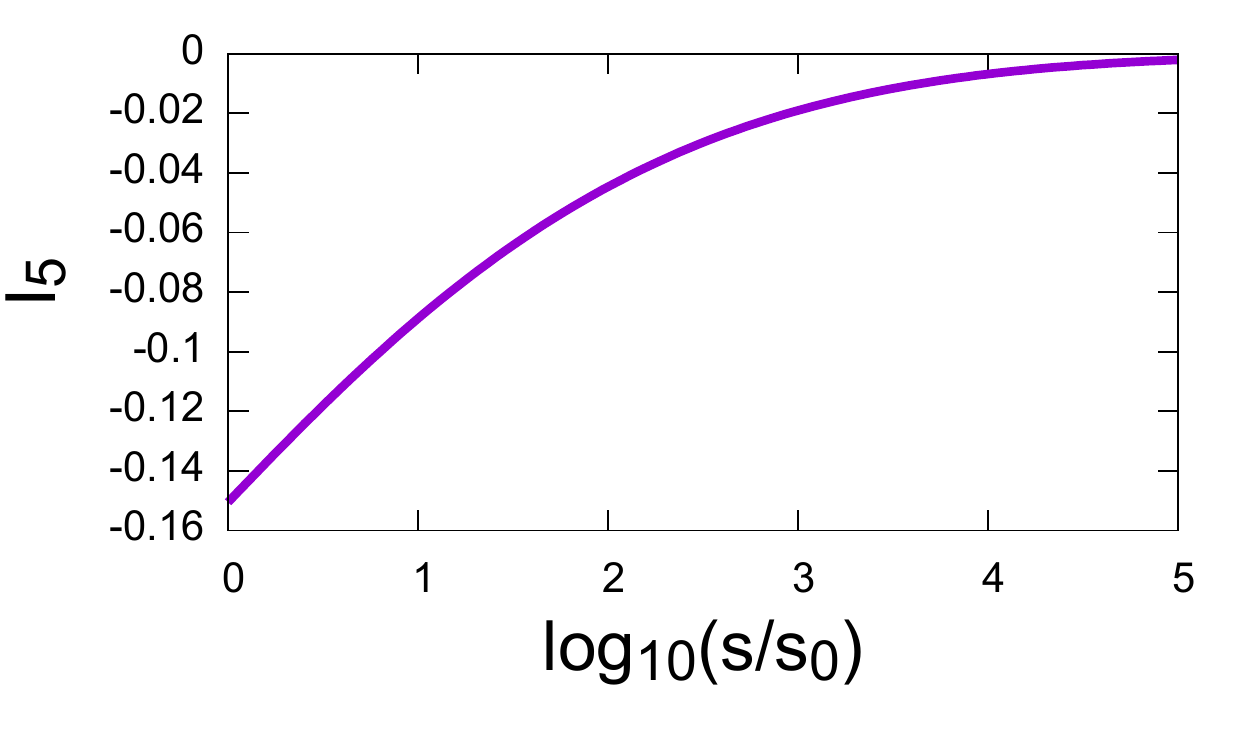}
\caption[]{The energy dependence of double-spin-flip $R_2$ (left) and single-spin-flip $r_5=R_5+iI_5$ (right and bottom) to non-flip ratios. 
Here, we take $\kappa = 1/16$ and $Q_{s0} = 1.0$ GeV.
}
\label{figImr2}
\end{center}
\end{figure}

The energy dependence of the total cross section computed in our approach is shown in Fig.~\ref{figcrosssec} with  $Q_{s0} = 1.0$ GeV. Here, the green and pink lines denote two different values of the starting energy scale, $\sqrt{s_0}=0.1$ and $0.5$ TeV, respectively. The result is in reasonable agreement with  the corresponding measurements in $pp$ collisions performed at several distinct energies, such as those by the TOTEM LHC Collaboration at 13 TeV \cite{Antchev:2017dia}, 8 TeV \cite{Antchev:2016vpy}, 7 TeV \cite{Antchev:2013iaa,Antchev:2013gaa} and 2.76 TeV \cite{Antchev:2018rec}, as well as in  $p\bar p$ collisions by D0 Tevatron Collaboration at 1.96 TeV \cite{Abazov:2012qb} and by UA4 CERN SPS Collaboration at 546 GeV \cite{Bernard:1987vq} and 630 GeV \cite{Bernard:1986ye}.  Since the measured values for $\sigma_{\rm tot}(s)$ are sometimes not available in the experimental articles, in those cases 
the $\sigma_{\rm tot}$ values have been taken from the global L\'evy analysis of the corresponding elastic $pp$ and $p\bar p$ cross section data performed recently in Ref.~\cite{Csorgo:2018uyp}. 
Incidentally, we have also tried $Q_{s0}=0.5$ GeV, but the quality of the fit is noticeably worse in this case. 

The results for $R_2$, $I_5$ and $R_5$ are plotted in Fig.~\ref{figImr2} as  functions of $\sqrt{s}$ in upper-left, upper-right and bottom panels, respectively. 
  Note that the normalization and sign of $R_{2,5}$ are arbitrary, as it is proportional to the unknown parameter $\kappa$, and we have chosen $R_5$  to be negative following the recent suggestion in \cite{Kopeliovich:2019yir}. Irrespective of this,  we can predict that $R_2$ and $R_5$ have the same sign and that $|R_2|$ is roughly two times larger than $|R_5|$.   We also see a clear tendency that the magnitude of $R_{2,5}$ decreases with increasing energy. This is because, although the Odderon intercept is unity in the dilute (BFKL) regime, the nonlinear saturation effect tends to suppress the Odderon amplitude \cite{Hatta:2005as,Lappi:2016gqe,Yao:2018vcg}.  On the other hand, the value of $I_5$ is a prediction of this model, since both the denominator and numerator of (\ref{eq:r5}) come from the Pomeron. It is negative and the magnitude decreases with energy  because of the factor $r_\perp^2$ in the numerator of  (\ref{eq:r5}): The $r_\perp$-integral is dominated by $r_\perp\sim 1/Q_s(s)$, and $Q_s(s)$ is an increasing function of energy.

The data on single spin asymmetry $A_N$ in small-angle elastic $pp$ collisions
\beq
A_N\frac{d\sigma}{dt} = -\frac{4\pi}{s(s-4M^2)}{\rm Im}\left\{(\phi_1 
+ \phi_2 + \phi_3 - \phi_4)\phi_5^*\right\} 
\,, \label{AN}
\eeq
have recently become available from the fixed-target measurement HJET at BNL \cite{Poblaguev:2019saw} as well as earlier from the STAR measurements of polarized elastic $pp$ collisions at $\sqrt{s}=200$ GeV \cite{Adamczyk:2012kn}. These data have enabled to extract the real and imaginary parts of $r_5$ ratio in a wide energy domain. The values of $R_5$ published by the experimental collaborations were found (by STAR measurement and by an extrapolation from the lower HJET energies) to be either small positive or consistent with zero at $\sqrt{s}=200$ GeV,
while the hadronic contribution predicted in Fig.~\ref{figImr2} (upper-right panel) is found 
to be larger than the ballpark of experimental values. 

We note, however, that the CNI contribution has to be taken into consideration as its impact 
on $r_5$ can be rather important, whereas the current analysis only focuses on the hadronic contribution to $\phi_5$. Indeed, as was recently advocated in Ref.~\cite{Kopeliovich:2019yir} relying on a Regge analysis and a dominance of the Pomeron spin-flip contribution, the absorptive corrections to the Coulomb spin-flip amplitude significantly modify the CNI mechanism. As a result, this modification affects the extracted values of $r_5$, in particular making the spin-flip Pomeron $I_5$ rather large and negative, at the level of $-5$ to $-10$ \% at $\sqrt{s}=200$ GeV, in consistency with expectations \cite{Kopeliovich:1989hp}. The fact that our QCD-based approach predicts non-vanishing and negative $I_5$ is encouraging, although as we explained above it falls with energy, in contrast to the behavior predicted by the Regge fit of Ref.~\cite{Kopeliovich:2019yir}. These results are not inconsistent and rather suggest that the gluon saturation regime has not been reached at RHIC energies. We leave a thorough analysis of the CNI effects in the current framework for a future work.

\section{Conclusions}
\label{Sec:conclusions}

In this work, we have presented a new QCD-inspired model for small-angle elastic proton-(anti)proton
scattering in terms spin-dependent Pomeron and Odderon helicity amplitudes in the dipole picture 
based upon the Wilson line approach. The elastic amplitudes $\phi_{1,\dots,5}$ are effectively 
described in near-forward kinematics by means of a scattering of the lowest Fock state 
$p\to q+(qq)$ of projectile proton (i.e.~the quark-diquark dipole) off the proton target, 
i.e.~in a similar fashion as DIS. The corresponding dipole $S$-matrix receives contributions 
from non-flip and spin-flip Pomeron and Odderon exchanges that are represented in terms 
of GTMDs of different types. 

Connecting to the numerical analysis of the small-$x$  Odderon evolution equation performed 
earlier in Ref.~\cite{Yao:2018vcg} and incorporating in addition the QCD running coupling effect, we explore
the relative importance of spin-flip contributions to the elastic $pp$ scattering at high energies.
In particular, we analyse the energy dependence of the spin-flip Pomeron ($I_5$) and spin-flip 
Odderon ($R_5$) amplitudes, as well as double-spin-flip Odderon ($R_2$) amplitude relative 
to the non-flip one. At variance with an earlier Regge-based calculation of Ref.~\cite{Kopeliovich:2019yir} 
incorporating for the first time the absorptive corrections in the CNI mechanism,
we do not assume that the exchanged spin-independent and spin-dependent Regge trajectories 
have different intercepts and do not neglect the Odderon contributions. Yet, we have reached a qualitatively 
similar conclusion about a significant and negative  contribution to the single helicity-flip   amplitude $I_5$. 
Moreover, the measured value of $R_5$ can be used to determine the Odderon coupling $\kappa$, which in turn determines the value of $R_2$. 
The energy dependence of $r_5$ 
in our approach is decaying and hence is strictly opposite to the steeply rising behavior from 
the Regge analysis \cite{Kopeliovich:2019yir} obtained in the lower energy region. This suggests that once 
the gluon saturation effect kicks in, the behavior of $r_5$ changes. A further analysis of this issue is certainly needed.

The experimentally probed energies in the existing measurements of the spin-flip contributions
may not be high enough to make a conclusive statement about the energy dependence of spin-dependent
Pomeron and, especially, Odderon effects. Indeed, at such low energies as $\sqrt{s}=200$ GeV the $C$-odd 
effects may come mostly from secondary Reggeon exchanges, not due to spin-dependent Odderon studied 
in our analysis here. It is therefore of high importance to perform a new measurement of $r_2$ and $r_5$ 
in a TeV energy range to make a definite conclusion about the energy dependence of spin-dependent Pomeron 
and Odderon in the future. Note that this does not necessarily require polarized proton beams which are not available at the LHC. The differential cross section (\ref{dt}) gets contributions from the helicity-flip amplitudes, but they are usually ignored in the CNI analysis. It would be very interesting to test more flexible parametrizations of the CNI effect including the hadronic and electromagnetic contributions to $\phi_{2,4,5}$. This could eventually affect the value of the $\rho$-parameter, and also the total cross section via   (\ref{corr}).

Finally, it is of course necessary to extend the present calculation to finite momentum transfer $t$, in particular up to the `dip' region of $d\sigma/dt$. The basic formulas are given in (\ref{phi1})-(\ref{phi5model}), but we are missing models of the spin-independent and spin-dependent Pomeron and Odderon amplitudes  at finite $t$ (see for example \cite{Dumitru:2018vpr,Dumitru:2020fdh} for a model of $g_{1,1}$ at finite impact parameter). They can also serve as an initial condition for the impact-parameter dependent BK equation to determine the energy dependence. It is also interesting to consider  different models for the 'slow' proton such as a bound state of three quarks. We hope to address these issues elsewhere.

\acknowledgments 
We thank Elliot Leader for useful correspondence. Y.~H. thanks Shandong University in QingDao, where this collaboration started, for hospitality.  
R.~P.~is supported in part by the Swedish Research Council grants, contract numbers 621-2013-4287 and 2016-05996, by the Ministry of Education, 
Youth and  Sports of the Czech Republic, project LTT17018, as well as by the European Research Council (ERC) under the European Union's Horizon 2020 
research and innovation programme (grant agreement No 668679). This work is supported by the U.S. Department of Energy, Office of Science, Office 
of Nuclear Physics, under contract No. DE- SC0012704, and in part by Laboratory Directed Research and Development (LDRD) funds from Brookhaven 
Science Associates.  J. Zhou has been supported by the National Science Foundations of
China under Grant No.\ 11675093, and by the Thousand Talents Plan
for Young Professionals.

\appendix

\section{One photon exchange}
In this Appendix we quickly reproduce the helicity amplitudes in the one-photon exchange approximation. For a complete result, see \cite{Buttimore:1978ry}.  
The scattering amplitude is given by 
\beq
iT&=&-e^2\bar{u}(P_3,S_3) \left(\gamma^\mu F_1 + \frac{i\sigma^{\mu\rho}\Delta_\rho}{2M}F_2\right)u(P_1,S_1) \nn && \qquad \times \frac{-i}{t}  
 \bar{u}(P_4,S_4) \left( \gamma_\mu F_1 - \frac{i\sigma_{\mu\lambda}\Delta^\lambda}{2M}F_2\right)u(P_2,S_2) \,, 
\eeq
where $\Delta= P_3-P_1 = P_2-P_4$ and $F_1$ and $F_2$ are Dirac and Pauli form factors. 
This immediately gives 
\beq
T_{++++} =T_{+-+-}=8\pi \phi_1= 4\pi \alpha_{em}\frac{2s}{t}F_1^2(t) \,.
\eeq
As for the double helicity-flip amplitudes, 
we use the formulas
\beq
\bar{u}_{-\lambda}(P_1) \sigma^{+i}\Delta^i u_\lambda(P_1) = 2i P_1^+\lambda \Delta_\perp \cdot \epsilon_\lambda, \qquad 
\bar{u}_{-\lambda}(P_2) \sigma^{-i}\Delta^i u_\lambda(P_2) = 2i P_2^-\lambda \Delta_\perp\cdot \epsilon^*_{\lambda},
\eeq
to get
\beq
\langle -\lambda,-\lambda'|T|\lambda,\lambda'\rangle  \approx 4\pi \alpha_{em}\frac{s}{2M^2} \frac{\lambda\lambda' \Delta_\perp \cdot \epsilon_\lambda \Delta_\perp \cdot \epsilon^*_{\lambda'} 
}{-t} F_2^2 .
\eeq
We therefore find  
\beq
T_{--++} = 8\pi \phi_2 = 4\pi \alpha_{em} \frac{s}{2M^2}F_2^2 .
\eeq
For $\phi_{4,5}$, we need to remove the phase according to (\ref{factor}). The results are  
\beq
T_{+--+}=
-4\pi \alpha_{em} \frac{s}{2M^2}\frac{(\Delta_\perp \cdot \epsilon_-)^2}{-t}F_2^2, \qquad \phi_4=-\alpha_{em}\frac{s}{4M^2}F_2^2 = -\phi_2.
\eeq
\beq
T_{+++-} =
4\pi \alpha_{em} \frac{s}{M} \frac{\Delta_\perp \cdot \epsilon_+}{-t} F_1F_2, \qquad \phi_5=-\alpha_{em}\frac{s}{2M\sqrt{-t}}F_1F_2.
\eeq

\section{Feynman rules of the quark-diquark model}

\begin{figure}[t]
\begin{center}
\includegraphics[width=11cm]{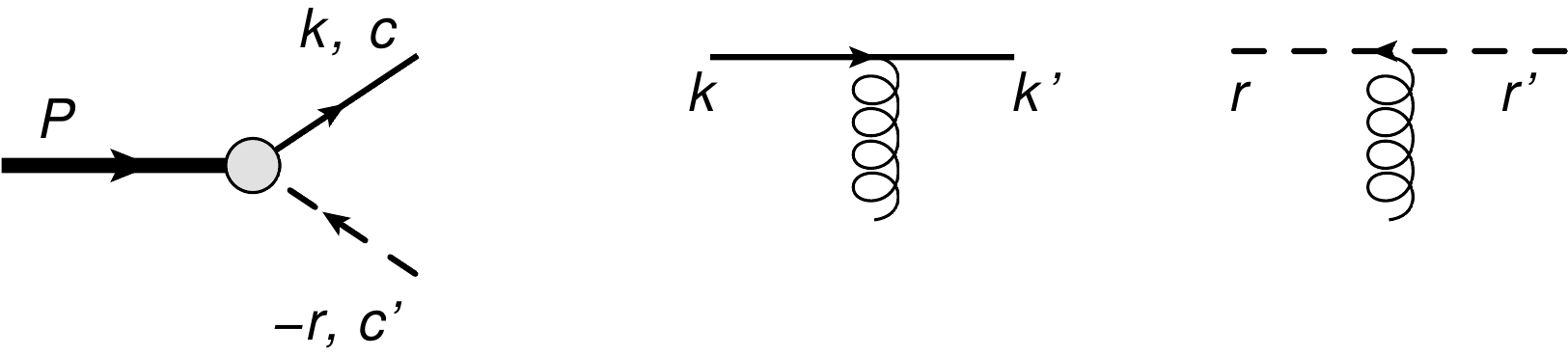}
\caption{Feynman rules of the quark-diquark model} \label{vertices}
\end{center}
\end{figure}

In the
diquark model \cite{Brodsky:2000ii}, the interaction between the nucleon, the quark, and the scalar diquark is described
by the following Feynman rules for the nucleon-quark-diquark vertex, quark-gluon vertex, and 
diquark-gluon vertex, respectively (see Fig.~\ref{vertices}),
\begin{eqnarray}
&& i c_s \bar u(k,\lambda_k)u(P,S_\bot)\delta^{cc'}, \ \ \,-igt^a\gamma^\mu,\ \ \ -igt^a(r+r')^\mu,
\end{eqnarray}
The  scalar diquark, quark and gluon propagators in the Feynman
gauge are given by 
\begin{equation}
\;\;\;\;\frac{i}{r^2-m_s^2+i\epsilon},\ \ \ \frac{i(\Slash k+m_q)}{k^2-m_q^2+i\epsilon}\;,\;\;\;\;
\frac{-i g^{\mu \nu}\delta^{c\,c'}}{k^2+i\epsilon}\;,
\label{prop}
\end{equation}
where  $c,c'$ are color indices in the adjoint representation
and $t^a$ 
are  SU(N) gauge group generators in the fundamental representation.

\bibliography{references}

\end {document}